\documentclass[twocolumn,secnumarabic,amssymb, nobibnotes, aps, prd]{revtex4-1}
\usepackage{mathrsfs}

\begin{document}

\title{Some generalizations of the Raychaudhuri equation}

\author{Gabriel Abreu}%
\email{gabriel.abreu@msor.vuw.ac.nz}
\affiliation{School of Mathematics, Statistics, and Operations Research; Victoria University of Wellington, Wellington, New Zealand.}
\author{Matt Visser}%
\email{matt.visser@msor.vuw.ac.nz}
\affiliation{School of Mathematics, Statistics, and Operations Research; Victoria University of Wellington, Wellington, New Zealand.}

\date{22 December 2010; \LaTeX-ed \today}%

\begin{abstract}
The Raychaudhuri equation has seen extensive use in general relativity, most notably in the development of various singularity theorems. 
In this rather technical article we shall generalize the Raychaudhuri equation in several ways. First an improved version of the standard timelike Raychaudhuri equation is developed, where several key terms are lumped together as a divergence.  This already has a number of interesting applications, both within the ADM formalism and elsewhere. Second, a spacelike version of the Raychaudhuri equation is briefly discussed.  Third, a version of the Raychaudhuri equation is developed that does not depend on the use of normalized congruences.  This leads to useful formulae for the ``diagonal'' part of the Ricci tensor. Fourth,  a ``two vector'' version of the Raychaudhuri equation is developed that uses two congruences to effectively extract ``off diagonal'' information concerning the Ricci tensor.

\bigskip
\noindent
Keywords:  Raychaudhuri  equation, timelike congruence, spacelike congruence, non-normalized congruence.


\end{abstract}
\pacs{}
\maketitle
\def\d{{\mathrm{d}}}
\newcommand{\scri}{\mathscr{I}}
\newcommand{\sun}{\ensuremath{\odot}}
\def\J{{\mathscr{J}}}
\def\sech{{\mathrm{sech}}}
\def\eof{\hfill$\Box$}
\def\implies{\Longrightarrow}
\def\tr{{\mathrm{tr}}}
\def\K{{{\textsc{K}}}}
\def\KK{{{\tiny\textsc{K}}}}

\section{Introduction}

The Raychaudhuri equation has become one of the standard workhorses of general relativity, particularly as applied to the singularity theorems. For textbook presentations see for instance~\cite{Hawking-Ellis, Wald, Carroll, Poisson}. Nevertheless, we feel that there are still some interesting ways in which the general formalism can be extended. 
There are four extended versions of the Raychaudhuri equation we wish to explore in this article:
\begin{itemize}
\item \emph{Single timelike unit vector field.}\\
By collecting several terms in the usual formulation into a divergence, we obtain a particularly useful version that finds many applications in the  ADM formalism and other situations.

\item \emph{Single spacelike unit vector field.}\\
This situation is most typically ignored. We will make a few hopefully clarifying comments.

\item \emph{Single non-normalized vector field.}\\
This somewhat simplifies  the Raychaudhuri equation, at the cost of no longer having nice positivity properties. 

\item \emph{Two non-normalized vector fields.}\\
This allows us to probe off-diagonal components of the Ricci tensor. 

\end{itemize}
These four extensions of the Raychaudhuri equation will soon be seen to each be useful in their own way, and to yield different information. We shall provide numerous examples below.

\section{Single unit timelike vector field}
This is the standard case. Let $u^a$ be a field of unit timelike vectors (a congruence).  This does not have to be the 4-velocity of a physical fluid (though it might be), it applies just as well to the 4-velocities of an imaginary collection of ``fiducial observers'' [FIDOs].
Then it is a purely geometrical result (see for example Hawking and Ellis~\cite{Hawking-Ellis}, pp 82--84, or Wald~\cite{Wald}, or Carroll~\cite{Carroll}, or Poisson~\cite{Poisson}, or even  {\sf Wikipedia}, (note that there are sometimes minor disagreements of notation --- typically just a factor of 2 in odd places) that:
\begin{equation}
\label{E:R0}
{\d\theta\over\d s} = - R_{ab} u^a u^b +  \omega^2 -  \sigma^2 - {1\over3}\theta^2 + \nabla_a \left( {\d u^a\over\d s} \right).
\end{equation}
\emph{This is the standard form of the Raychaudhuri equation.}
The spatial projection tensor is
\begin{equation}
h_{ab} = g_{ab} + u_a u_b.
\end{equation}
This projection tensor has signature $\{0,+1,+1,+1\}$.
Various shear and expansion related quantities are 
\begin{equation}
\theta_{ab} = h_{ac} \nabla^{(c} u^{d)} h_{db};
\end{equation}
\begin{equation}
\theta=  g^{ab}\theta_{ab} = h^{ab} \theta_{ab} = \nabla_a u^a;
\end{equation}
\begin{equation}
\sigma_{ab} = \theta_{ab} - {1\over3} h_{ab} \theta;
\end{equation}
\begin{equation}
\sigma^2 =  \sigma_{ab} \sigma^{ab} \geq 0.
\end{equation}
Vorticity related quantities are
\begin{equation}
\omega_{ab} = h_{ac} \nabla^{[c} u^{d]} h_{db};
\end{equation}
\begin{equation}
\omega^2 = \omega_{ab} \omega^{ab} \geq 0.
\end{equation}
With these definitions we have the usual decomposition
\begin{equation}
u_{a;b} = \omega_{ab} + \sigma_{ab} +{1\over3} \theta h_{ab} - {\d u_a\over\d s} \, u_b.
\end{equation}
See (for example) pages 82--84 of Hawking and Ellis~\cite{Hawking-Ellis}.
Equation (\ref{E:R0})  is Wald's equation (9.2.11)~\cite{Wald}, supplemented with the $\nabla_a \left( {\d u^a\over\d s} \right)$ term due to allowing a non-geodesic congruence; you can deduce the presence of this term from the second line in his (9.2.10) by not assuming geodesic motion.

Now consider the identity
\begin{equation}
{\d \theta\over \d s} = u \cdot \nabla \theta = \nabla\cdot (\theta u) - \theta \nabla\cdot u =  \nabla\cdot (\theta u) - \theta^2.
\end{equation}
Using this identity we can also write the Raychaudhuri equation in the slightly unusual forms
\begin{equation}
\label{E:R1}
\nabla_a \left( \theta u^a - {\d u^a\over\d s} \right)= - R_{ab} u^a u^b +  \omega^2 -  \sigma^2 + {2\over3}\theta^2 ,
\end{equation}
or
\begin{equation}
\label{E:R2}
R_{ab} u^a u^b= \omega^2 -  \sigma^2 + {2\over3}\theta^2 +  \nabla_a \left(- \theta \; u^a + {\d u^a\over\d s} \right). 
\end{equation}
This extension of the usual Raychaudhuri equation is ``close'' to, but significantly more general than,  a key technical result used by Padmanabhan and Patel in references~\cite{paddy1, paddy2, paddy3}. 

\section{Applications}
We now consider several applications of the above formalism --- these applications basically amount  to strategically choosing an appropriate congruence.

\subsection{Vorticity-free congruence}
\label{S:vorticity-free}

Let $\Psi(x)$ be an arbitrary scalar field and define a set of fiducial observers [FIDOs] by
\begin{equation}
u_a \propto \nabla_a \Psi.
\end{equation}
Then normalizing we have
\begin{equation}
u_a = - {\nabla_a \Psi\over ||\nabla \Psi||},
\end{equation}
and furthermore
\begin{equation}
\omega_{ab}=0.
\end{equation}
The minus sign here is purely conventional, it guarantees that the $u^a$ is ``future-pointing'' in the direction of increasing $\Psi$.
Conversely
\begin{equation}
\omega_{ab}=0 \qquad \implies \qquad u_a \propto \nabla_a \Psi.
\end{equation}
This is guaranteed by the Frobenius theorem.

Then in this vorticity-free situation the extended Raychaudhuri equation reduces to
\begin{equation}
{\d\theta\over\d s} = - R_{ab} u^a u^b  - \sigma^2 - {1\over3}\theta^2 + \nabla_a \left( {\d u^a\over\d s} \right),
\end{equation}
or equivalently
\begin{equation}
\label{E:18}
R_{ab} u^a u^b=  -  \sigma^2 + {2\over3}\theta^2 +  \nabla_a \left(- \theta u^a + {\d u^a\over\d s} \right),
\end{equation}
or even
\begin{equation}
R_{ab} u^a u^b=  -  \theta_{ab} \theta^{ab}  +  \theta^2 + \nabla_a \left(- \theta u^a + {\d u^a\over\d s} \right).
\end{equation}
But since $u^a$ is now hyper-surface orthogonal we can use the slices of constant $\Psi$ to define a spacelike foliation --- the scalar $\Psi$ serves (at least locally)  as a ``cosmic time'' function.  Then in terms of the extrinsic curvature $K_{ab}$ of the constant $\Psi$ hyper-surfaces we have, (using Misner, Thorne, and Wheeler~\cite{MTW} sign conventions for the extrinsic curvature), the results:
\begin{equation}
\theta_{ab} =  -K_{ab}; \qquad \theta =  - K;
\end{equation}
\begin{equation}
\sigma_{ab} =  - \left( K_{ab} - {1\over3} K h_{ab} \right);
\end{equation}
\begin{equation}
\sigma^2 = {1\over2} \left[ K_{ab} K^{ab} -{1\over3} K^2 \right].
\end{equation}
But then
\begin{equation}
\label{E:paddy1}
R_{ab} u^a u^b=  - K_{ab} K^{ab}  + K^2 +  \nabla_a \left( K\; u^a + {\d u^a\over\d s} \right).
\end{equation}
This is effectively one of the key technical results used by Padmanabhan and Patel in~\cite{paddy1, paddy2, paddy3}, but now we see that this result is actually a special case of a considerably more general result, and can be viewed as a relatively straightforward extension and then specialization of the Raychaudhuri equation.

\subsection{ADM formalism}

By definition, in any stably causal spacetime there is a globally defined  ``cosmic time'' function $t(x)$ such that $\d t$ is always timelike. Then on the one hand the constant-$t$ slices are always spacelike and can be used to set up an ADM decomposition of the metric, while on the other hand $u = - (\d t)^\sharp /||\d t||$ is a vorticity-free unit timelike congruence, so that the results of the previous subsection apply.  (As usual, $\d t^\sharp$ denotes the vector obtained form the one-form $\d t$ by ``raising the index'', similarly $u^\flat$ will denote the one-form obtained from the vector $u$ by ``lowering the index.) 

Consequently the extended Raychaudhuri equation can now be cast in the form
\begin{equation}
\label{E:paddy2}
R_{\hat t \hat t} =  - K_{ab} K^{ab}  + K^2 +  \nabla_a \left( K\; u^a + {\d u^a\over\d s} \right).
\end{equation}
This result complements and reinforces the information one obtains from the Gauss equations --- see for example Misner, Thorne, and Wheeler~\cite{MTW} pp 505--520, or Rendall~\cite{Rendall} pp 23--24.   The Gauss equations (for a spacelike hypersurface) are
\begin{equation}
^{(4)} R_{abcd} = {}^{(3)} R_{abcd}  + K_{ac} K_{bd} - K_{ad} K_{bc}.
\end{equation}
Contracting once
\begin{equation}
^{(4)} R_{ab} =  {}^{(3)}R_{ab} -  {}^{(4)}R_{acbd} u^c u^d + \tr(K) K_{ab} - (K^2)_{ab}.
\end{equation}
Contracting a second time
\begin{equation}
^{(4)} R =  {}^{(3)}R - 2 \, {}^{(4)}R_{ab} u^a u^b + K^2 - \tr(K^2).
\end{equation}
But now, since ${}^{(4)}R_{ab} u^a u^b$ has been given to us via the extended Raychaudhuri equation, we easily see that for a spacelike hypersurface
\begin{equation}
^{(4)} R =  {}^{(3)}R + \tr(K^2) - K^2 -2 \, \nabla_a \left( K\; u^a + {\d u^a\over\d s} \right).
\end{equation}
Traditional derivations of this result are sometimes somewhat less transparent, and viewing it as an extension of the timelike Raychaudhuri equation is the cleanest derivation we have been able to develop. 
To see some of the deeper connections with the ADM formalism read (for example) \S 21.6 on pp 519--520 of Misner, Thorne, and Wheeler~\cite{MTW}; note especially eq (21.88).  See also exercise (21.10) on p 519. Also note the discussion by by Padmanabhan and Patel in references~\cite{paddy1, paddy2, paddy3}. 
Also, we should warn the reader that Wald uses an opposite sign convention for the extrinsic curvature. See specifically Wald~\cite{Wald} equation (10.2.13) on page 256.

\subsection{Static spacetimes}

Let us now take the discussion in a rather different direction, and assume that the spacetime is {\emph{static}}. That is, there exists a hypersurface-orthogonal Killing vector $k^a$ that is timelike at spatial infinity. Because it is hypersurface orthogonal then $k_a \propto \nabla_a \Psi$, and so $u^a = k^a/||k||$ is a set of FIDOs of the type considered in the  previous section.  But since $k^a$ is also a Killing vector we have $k_{(a;b)}=0$ and so obtain the quite standard result that
\begin{eqnarray}
\label{E:killing-shear}
u_{(a;b)} &=&  \nabla_{(a} \{k/||k||\}_{b)} = {k_{(a;b)}\over||k||} - {k_{(b} \nabla_{a)} ||k||\over ||k||^2} 
\nonumber\\
&=&  - {k_{(b} \nabla_{a)} ||k||\over ||k||^2} 
 =  - {u_{(b} \nabla_{a)} ||k||\over ||k||} 
 \nonumber\\
&=&  - {u_{(b}  ||k||_{,a)}\over ||k||}  =   {u_{(a}  ||k||_{,b)}\over ||k||}.
\end{eqnarray}
Hence
\begin{equation}
\theta_{ab}=0 \quad\implies\quad  K_{ab}=0  \quad\implies\quad  K=0.
\end{equation}
That is, in static spacetimes the extrinsic curvature of the time-slices is zero (in addition to the congruence being vorticity free). The Raychaudhuri equation then specializes to the particularly simple result
\begin{equation}
R_{ab} \; u^a u^b= \nabla_a \left( {\d u^a\over\d s} \right).
\end{equation}
This is essentially the technical result we used in our derivation of an entropy bound for static spacetimes~\cite{Abreu1, Abreu2}, though in those articles we had derived it from an old result due to Landau and Lifshitz~\cite{Landau}. (The original Landau--Lifshitz result is obtained via a straightforward but tedious series of index manipulations, with little geometrical insight.) 

\subsection{Stationary spacetime --- Killing congruence}
\label{S:Killing}

What can we now do for \emph{stationary}, as opposed to \emph{static} spacetimes? (This distinction is relevant to ``rotating spacetimes'', for example Kerr spacetimes versus Schwarzschild spacetimes. See for instance~\cite{Kerr1, Kerr2, Kerr-book, Kerr-survey}.)  The (asymptotically) timelike Killing vector $k= \partial_t$  [that is, $k^a = (1; 0,0,0)^a$] is no longer hypersurface orthogonal.  Nevertheless we can still define the timelike Killing congruence
\begin{equation}
u^a =  {k^a\over ||k||}.
\end{equation}
This timelike congruence corresponds to a class of FIDOs [not ZAMOs, not zero angular momentum observers] that sit at fixed spatial coordinate position~\cite{membrane, Abreu3}.
This timelike congruence, even though it is \emph{not} hypersurface orthogonal, still satisfies equation (\ref{E:killing-shear}).
So even though there is no longer any interpretation of the shear in terms of an extrinsic curvature, we still have
\begin{equation}
\theta_{ab}=0,
\end{equation}
whence both
\begin{equation}
 \sigma_{ab}=0;  \qquad\hbox{and} \qquad  \theta=0.
\end{equation}
Therefore
\begin{equation}
R_{ab} u^a u^b= \omega^2 +  \nabla_a \left( {\d u^a\over\d s} \right). 
\end{equation}
However, unless further assumptions are made, we cannot do much with the $\omega^2$ term. 
Generically we have
\begin{eqnarray}
u_{[a;b]} &=&  \nabla_{[a} \{k/||k||\}_{b]} = {k_{[a;b]}\over||k||} - {k_{[b} \nabla_{a]} ||k||\over ||k||^2} 
\nonumber\\
&=&   {k_{[a;b]}\over||k||}  - {k_{[b} \nabla_{a]} ||k||\over ||k||^2} 
=  {k_{[a;b]}\over||k||}   - {u_{[b} \nabla_{a]} ||k||\over ||k||} 
\nonumber\\
&=&  {k_{[a;b]}\over||k||}   - {u_{[b}  ||k||_{,a]}\over ||k||}.
\end{eqnarray}
This implies
\begin{equation}
\omega^{ab} = h^{ac} h^{bd} \; {k_{[c;d]}\over||k||},
\end{equation}
whence
\begin{equation}
R_{ab} u^a u^b= +{h^{ac} h^{bd} \;  k_{[a;b]} \; k_{[c;d]}\over||k||^2 }  +  \nabla_a \left( {\d u^a\over\d s} \right).
\end{equation}
Unfortunately this does not simplify any further, and without further assumptions for the timelike Killing congruence on a stationary spacetime we should just be satisfied by the \emph{inequality}:
\begin{equation}
\label{E:inequality}
R_{ab} u^a u^b \geq  \nabla_a \left( {\d u^a\over\d s} \right).
\end{equation}

\subsection{Stationary axisymmetric spacetimes}

In a stationary axisymmetric spacetime consider the vorticity-free congruence of section~\ref{S:vorticity-free} (not the Killing congruence of section~\ref{S:Killing}).  Because of the axisymmetry the congruence $u = - (\d t)^\sharp/||\d t||$ must then be a linear combination of the two Killing vectors, $k_t=\partial_t$ and $k_\phi=\partial_\phi$, in which case $\theta=\nabla\cdot u = 0$. In this case equation (\ref{E:18}) reduces to
\begin{equation}
\label{E:18b}
R_{ab} u^a u^b=  -  \sigma^2 +  \nabla_a \left( {\d u^a\over\d s} \right),
\end{equation}
which implies, for the natural vorticity-free congruence on an stationary axisymmetric  spacetime 
\begin{equation}
\label{E:18c}
R_{ab} u^a u^b \leq  \nabla_a \left( {\d u^a\over\d s} \right).
\end{equation}
It is this particular inequality that we used in reference~\cite{Abreu3} to place an entropy bound on rotating fluid blobs. 
(Note that the direction of the inequality has changed between equations (\ref{E:inequality}) and (\ref{E:18c}), but that is merely due to the fact that we are using different timelike congruences.)

\section{Single unit spacelike vector field}

In counterpoint, we now let $u^a$ be a field of unit spacelike vectors. The projection tensor becomes
\begin{equation}
h_{ab} = g_{ab} - u_a u_b.
\end{equation}
In contrast to the timelike situation the projection tensor is now of indefinite signature $\{-1,+1,+1,0\}$. One can still formally define the quantities $\theta_{ab}$, $\theta$, $\sigma_{ab}$, and $\omega_{ab}$, but they no longer have the same physical interpretation in terms of shear and vorticity. Furthermore since the projection tensor has indefinite signature we now \emph{cannot} guarantee either $\sigma^2\geq0$ or $\omega^2 \geq 0$.  On the other hand, the Raychaudhuri equation is formally unaffected.  That is, the fundamental equations (\ref{E:R0}), (\ref{E:R1}), and (\ref{E:R2}),  continue to hold as they stand. 

If we now consider a vorticity-free spacelike congruence, it will be hypersurface orthogonal to a timelike hypersurface. (That is, the normal to the hypersurface is spacelike, while the tangent space to the hypersurface can be chosen to have a basis of one timelike and two spacelike tangent vectors.) 

In this situation  we can without loss of generality set $u = (\d\Psi)^\sharp/||\d\Psi||$. Then $\omega_{ab}\to 0$, while in terms of the extrinsic curvature  $\sigma_{ab} \to -K_{ab}$ as for vorticity free timelike congruencies.  Thus equation (\ref{E:paddy1}) is formally unaffected and can now be cast in the form
\begin{equation}
\label{E:paddy3}
R_{\hat n \hat n} =  - K_{ab} K^{ab}  + K^2 +  \nabla_a \left( K\; u^a + {\d u^a\over\d s} \right).
\end{equation}

On the other hand, because $u$ is now a spacelike normal to a timelike hypersurface there is a key sign flip in the Gauss equations, which now read
\begin{equation}
^{(4)} R_{abcd} = {}^{(3)} R_{abcd}   - K_{ac} K_{bd} +K_{ad} K_{bc}  .
\end{equation}
Contracting twice
\begin{equation}
^{(4)} R =  {}^{(3)}R + 2 \, {}^{(4)}R_{ab} u^a u^b + \tr(K^2) - K^2.
\end{equation}
Therefore for a timelike hypersurface we have
\begin{equation}
^{(4)} R =  {}^{(3)}R - \tr(K^2) + K^2 +2 \, \nabla_a \left( K\; u^a + {\d u^a\over\d s} \right).
\end{equation}

In summary, for spacelike congruences the Raychaudhuri equation itself is formally unaffected (though the projection tensor is slightly different and we can no longer rely on the non-negativity of $\sigma^2$ and $\omega^2$). However applications of the Raychaudhuri equation, specifically anything involving the Gauss equations for embedded hypersurfaces, typically exhibit a limited number of sign flips.

\section{Single non-normalized vector field}

Now consider an \emph{non-normalized} vector field $u^a$, either spacelike, timelike, or null. What if anything can we say about the quantity
\begin{equation}
R_{ab} u^a u^b = \quad ???
\end{equation}
Following and modifying the discussion of Wald~\cite{Wald}, see (E.2.28) on page 464:
\begin{eqnarray}
R_{ab} u^a u^b &=& R^c{}_{acb} u^a u^b 
\nonumber\\
&=& - u^a \left[ \nabla_a \nabla_b - \nabla_b \nabla_a \right] u^b
\nonumber\\
&=& - \nabla_a (u^a \nabla_b u^b)  + (\nabla_a u^a)(\nabla_b u^b) 
\nonumber\\
&& + \nabla_b (u^a \nabla_a u^b) - (\nabla_b u^a)(\nabla_a u^b) \qquad
\nonumber\\
&=& \nabla_a (- u^a \nabla_b u^b + u^b \nabla_b u^a)  
\nonumber\\
&&+ (\nabla\cdot u)^2 - (\nabla_b u_a)(\nabla^a u^b)  
\nonumber\\
&=& \nabla\cdot\{ (u\cdot \nabla) u - (\nabla\cdot u) u \} 
\nonumber\\
&&+ (\nabla\cdot u)^2 - (\nabla_b u_a)(\nabla^a u^b) .
\end{eqnarray}
In obvious notation, using $\theta=\nabla\cdot u$, this can be cast as
\begin{eqnarray}
R_{ab} u^a u^b 
&=& \nabla\cdot\{ \nabla_u u - \theta u \} + \theta^2 
\nonumber\\
&& - \nabla_{(a} u_{b)} \nabla^{(a} u^{b)}    + \nabla_{[a} u_{b]} \nabla^{[a} u^{b]}.
\end{eqnarray}
This result can be viewed as another generalization of the Raychaudhuri equation.
The advantage of this particular formula is that we have not carried out any projections, and have not even committed ourselves to the nature of the congruence, be it spacelike, timelike, or null.  One disadvantage is that because of the Lorentzian signature of spacetime we \emph{cannot} (at least not without further assumption) guarantee
\begin{eqnarray}
\nabla_{(a} u_{b)} \; \nabla^{(a} u^{b)}   \geq 0 \quad???
\end{eqnarray}
\begin{eqnarray}
 \nabla_{[a} u_{b]} \; \nabla^{[a} u^{b]}   \geq 0 \quad ???
\end{eqnarray}
Two specific applications come readily to mind:
\begin{itemize}

\item 
For any Killing vector $u^a = k^a$ we have $\nabla_{(a} u_{b)}=0$, and consequently $\theta=0$. Therefore for any Killing vector whatsoever
\begin{eqnarray}
R_{ab} k^a k^b 
&=& \nabla\cdot\{ \nabla_k k  \}  + \nabla_{[a} k_{b]} \nabla^{[a} k^{b]} .  
\end{eqnarray}

\item
For any one arbitrary exact one-form $u = d \Psi$, even a locally exact one-form, we have $\nabla_{[a} u_{b]}=0$, while $\theta=\nabla^2\Psi$ and $\nabla_{(a} u_{b)} \nabla^{(a} u^{b)} = \Psi_{;a;b} \Psi^{;a;b}$. Therefore for any locally exact one-form whatsoever
\begin{eqnarray}
R^{ab} (\d \Psi)_a (\d \Psi)_b 
&=& \nabla\cdot\{ \nabla_{\d\Psi}\d\Psi - (\nabla^2\;\Psi)  \d\Psi \} 
\nonumber\\
&&+ (\nabla^2\Psi)^2  -  \Psi_{;a;b} \,\Psi^{;a;b}.
\end{eqnarray}
In fact, $\Psi$ could simply be one of the spacetime coordinates (defined on some suitable local coordinate patch) in which case this version of the Raychaudhuri equation turns into a statement about the diagonal components of the Ricci tensor in a coordinate basis
\begin{eqnarray}
R^{\Psi\Psi}
&=& \nabla\cdot\{ \nabla_{\d\Psi}\d\Psi - (\nabla^2\;\Psi)  \d\Psi \} 
\nonumber\\
&&+ (\nabla^2\Psi)^2  -  \Psi_{;a;b} \,\Psi^{;a;b}.
\end{eqnarray}
More boldly, if one chooses $\Psi$ to be a harmonic coordinate, ($\nabla^2 \Psi=0$), and this can always be done locally, then
\begin{eqnarray}
R^{\Psi\Psi}
&=& \nabla\cdot\{ \nabla_{\d\Psi}\d\Psi  \} 
  -  \Psi_{;a;b} \,\Psi^{;a;b}.
\end{eqnarray}
\end{itemize}
In summary, this extension of the Raychaudhuri equation has given us some useful computational formulae.

\section{Two non-normalized vector fields}

We shall now ask if it is possible to extract any useful information by considering two different congruences simultaneously. 

\subsection{Motivation}

To motivate this particular extension of the Raychaudhuri equation, recall that many decades ago Landau and Lifshitz had shown that in any stationary spacetime~\cite{Landau} (\S 105 equation (105.22),  for a recent application of this result see~\cite{Abreu1, Abreu2}):
\begin{equation}
R_{0}{}^{0} = {1\over\sqrt{-g_4}} \, \partial_i \left( \sqrt{-g_4} \, g^{0a} \, \Gamma^i{}_{a0} \right).
\end{equation}
(Here $a\in\{0,1,2,3\}$; $i\in\{1,2,3\}$.) But because the metric is stationary ($t$ independent) we can also write this as
\begin{equation}
R_{0}{}^{0} ={1\over\sqrt{-g_4}} \, \partial_b \left( \sqrt{-g_4} \, g^{0a} \, \Gamma^b{}_{a0} \right).
\end{equation}
To begin converting this into a coordinate-free statement, note that
\begin{equation}
R_{0}{}^{0} = R^a{}_b (\d t)_a (\partial_t)^b =  R^{a}{}_{b} \; (\d t)_a \; k^b.
\end{equation} 
Here we have had to use \emph{both} the timelike Killing vector $k$, for which $k^a = (\partial_t)^a =(1,0,0,0)^a$, \emph{and} the one-form  $\d t$, for which $(\d t)_a= (1,0,0,0)_a$.
By direct computation
\begin{eqnarray}
 g^{0a} \, \Gamma^b{}_{a0}  &=&   g^{ca} \, \Gamma^b{}_{ad} \; (\d t)_c  k^d  = \Gamma^b{}_{cd} \; (\d t)^c k^d   
 \nonumber\\
 &=& \Gamma^b{}_{cd} \; k^c  (\d t)^d 
= \{\partial_d  k^b +\Gamma^b{}_{cd} k^c \} \,  (\d t)^d
 \nonumber\\
&=&  (\nabla_d k^b) (\d t)^d
= (\d t)^d (\nabla_d k^b).
\end{eqnarray}
But then
\begin{eqnarray}
R_0{}^0 &=& 
{1\over\sqrt{-g_4}} \, \partial_b \left( \sqrt{-g_4} \, g^{0a} \, \Gamma^b{}_{a0} \right)
\nonumber\\
&=& {1\over\sqrt{-g_4}} \, \partial_b \left( \sqrt{-g_4}  (\d t)^d (\nabla_d k^b)  \right)
\nonumber\\
&=&  \nabla_b \{ (\d t)^d (\nabla_d k^b) \}.
\end{eqnarray}
So  the Landau--Lifshitz result is equivalent to the statement that in any stationary spacetime
\begin{equation}
R^{a}{}_{b} \; (\d t)_a k^b = \nabla_b \{  (\d t)^d (\nabla_d k^b) \} = \nabla \cdot ( \nabla_{\d t^\sharp} k ).
\end{equation}
So some linear combination of Ricci tensor components is given by a pure divergence.  Note that two \emph{different} vector fields are involved. This observation naturally leads to the question: Is it possible to come up with a variant of the Raychaudhuri equation that depends on \emph{two} congruences $u^a$ and $v^a$? Something of the form
\begin{equation}
R_{ab} \; u^a v^b = \quad ???
\end{equation}
We shall see how this is done below.
 
For now, let us mention that
\begin{eqnarray}
(\nabla_d k^b) (\d t)^d &=&  (\nabla^d k^b) (\d t)_d = -  (\nabla^b k^d) (\d t)_d 
\nonumber\\
&=& - \nabla^b \{k^d (\d t)_d\} + k^d \nabla^b (\d t)_d 
\nonumber\\
&=&   - \nabla^b \{ 1\} + k^d \nabla^b \nabla_d t 
\nonumber\\
&=&  k^d \nabla^b \nabla_d t = k^d \nabla_d \nabla^b t.
\end{eqnarray}
So the Landau--Lifshitz  result can also be written in the alternative form
\begin{equation}
R^{a}{}_{b} \; (\d t)_a k^b = \nabla_b \{ k^d \nabla_d \nabla^b t   \} = \nabla\cdot(\nabla_k \d t^\sharp).
\end{equation}
Finally note that
\begin{equation}
(\nabla_d k^b) (\d t)^d (\d t)_b=  (\nabla_d k_b) (\d t)^d (\d t)^b = 0, 
\end{equation}
so the vector $\nabla_{\d t^\sharp} k= \nabla_k \d t^\sharp$ is perpendicular to $\d t^\sharp$.

\subsection{Construction}

Following and modifying the discussion of Wald~\cite{Wald}, see equation  (E.2.28) on page 464:
\begin{eqnarray}
R_{ab} u^a v^b 
&=& 
R^c{}_{acb} u^a v^b 
\nonumber\\
&=& 
- u^a \left[ \nabla_a \nabla_b - \nabla_b \nabla_a \right] v^b
\nonumber\\
&=& 
- \nabla_a (u^a \nabla_b v^b)  + (\nabla_a u^a)(\nabla_b v^b) 
\nonumber\\
&&
+ \nabla_b (u^a \nabla_a v^b) - (\nabla_b u^a)(\nabla_a v^b)\qquad
\nonumber\\
&=& 
\nabla_a (- u^a \nabla_b v^b + u^b \nabla_b v^a)  
\nonumber\\
&&
+ (\nabla\cdot u) (\nabla\cdot v) - (\nabla_b u_a)(\nabla^a v^b).
\end{eqnarray}
With minor notational changes and given the symmetry of the Ricci tensor this can also be written as
\begin{eqnarray}
R_{ab} u^a v^b 
&=& \nabla\cdot\{ (u\cdot \nabla) v - (\nabla\cdot v) u \} 
\nonumber\\
&&+ (\nabla\cdot u) (\nabla\cdot v) - (\nabla_b u_a)(\nabla^a v^b),
\end{eqnarray}
and
\begin{eqnarray}
R_{ab} u^a v^b 
&=& \nabla\cdot\{ (v\cdot \nabla) u - (\nabla\cdot u) v \} 
\nonumber\\
&&+ (\nabla\cdot u) (\nabla\cdot v) - (\nabla_b u_a)(\nabla^a v^b) .
\end{eqnarray}
Furthermore (in obvious notation) this can again be rewritten as
\begin{eqnarray}
R_{ab} u^a v^b 
&=& \nabla\cdot\{ \nabla_u v - \theta_v u \} + \theta_u \theta_v 
\nonumber\\
&&- \nabla_{(a} u_{b)} \nabla^{(a} v^{b)}    + \nabla_{[a} u_{b]} \nabla^{[a} v^{b]}.
\label{E:crossed1}
\end{eqnarray}
and
\begin{eqnarray}
R_{ab} u^a v^b 
&=& \nabla\cdot\{ \nabla_v u - \theta_u v \} + \theta_u \theta_v 
\nonumber\\
&&- \nabla_{(a} u_{b)} \nabla^{(a} v^{b)}    + \nabla_{[a} u_{b]} \nabla^{[a} v^{b]}. 
\label{E:crossed2}
\end{eqnarray}
Note the similarities to the single-congruence case,  and note particularly the presence of a divergence term generalizing the standard Raychaudhuri equation. 
To check the equivalence of these two formulae note
\begin{eqnarray}
(\nabla_u v - \theta_v u) &-& (\nabla_ v u- \theta_u v) 
\nonumber
\\
&=&  [\nabla_u v +\theta_u v] - [\nabla_v u +\theta_v u] 
\nonumber\\
&=& \nabla\cdot[u\otimes v - v\otimes u]
\nonumber\\
&=& \nabla\cdot[u\wedge v].
\end{eqnarray}
That is, the difference of these two currents is the divergence of a 2-form, which makes it automatically closed.

\subsection{Generalizing the Landau--Lifshitz result}

Let $u=k$ be any Killing vector, and let  $v^\flat$ be an arbitrary (locally) exact one-form, so $v= (\d\Psi)^\sharp$ where $\Psi(x)$ is an arbitrary scalar. Then $ \nabla_{(a} u_{b)} =  \nabla_{(a} k_{b)}  = 0$, and so $\theta_u=0$. Furthermore $\nabla_{[a} v_{b]} = \nabla_{[a} \nabla_{b]} \Psi = 0$, so from equation (\ref{E:crossed1}) we have
\begin{equation}
R_{ab} \; k^a \nabla^b \Psi = \nabla\cdot \{   \nabla_k \d\Psi -  (\nabla^2\Psi) k\},
\end{equation}
while from equation (\ref{E:crossed2}) we have
\begin{equation}
R_{ab} \; k^a \nabla^b \Psi = \nabla\cdot \{   \nabla_{\d\Psi^\sharp}  k\} .
\end{equation}
This nicely generalizes the Landau--Lifshitz result to any arbitrary Killing vector and any arbitrary (locally) exact one form $\d\Psi$, not just $\d t$.
 (That these two formulae are equivalent follows from the discussion in the previous section above.)   
Note the (standard) Landau--Lifshitz result corresponds to $k^a\to(\partial_t)^a$ and $\Psi \to t$.

Now choose a coordinate system adapted to the Killing vector $k$. Let $k = \partial_{\KK}$ define a Killing coordinate $\K$, so that all geometrical objects are independent of the coordinate $\K$. Let $\Psi$ also be viewed as a coordinate, relabel it as $x^a$, possibly distinct from $\K$, and with no claim that $x^a$ necessarily corresponds to a Killing vector.  Then
\begin{equation}
R_{\KK}{}^a = \nabla\cdot \{   \nabla_{(\d x^a)^\sharp}  \partial_{\KK}\} .
\end{equation}
Unwrapping the covariant derivatives we see
\begin{equation}
R_{\KK}{}^{a} ={1\over\sqrt{-g_4}} \, \partial_b \left( \sqrt{-g_4} \, g^{a c} \, \Gamma^b{}_{c\KK} \right).
\end{equation}
If we now let the index $i$ range over every coordinate except the Killing coordinate $\K$ then, because all geometrical objects are independent of the coordinate $\K$, we have
\begin{equation}
R_{\KK}{}^{a} ={1\over\sqrt{-g_4}} \, \partial_i \left( \sqrt{-g_4} \, g^{a c} \, \Gamma^i{}_{c\KK} \right).
\end{equation}
This equation, ultimately based on our two-congurence extension of the Raychaudhuri equation (\ref{E:crossed1}), is now very much in Landau--Lifshitz form, but us definitely considerably more powerful than the original Landau--Lifshitz result.

\subsection{Landau--Lifshitz in axial symmetry}

Since in a stationary spacetime with axial symmetry we have a second azimuthal Killing vector $k^a\to (\partial_\phi)^a$, 
and could also consider $\Psi \to \phi$, then there are three additional Landau--Lifshitz like results:
\begin{equation}
R_\phi{}^t = R_{ab} (\partial_\phi) ^a \nabla^b t = \nabla\cdot \{   \nabla_{\d t^\sharp}  \partial_\phi \};
\end{equation}
\begin{equation}
R_t{}^\phi = R_{ab} (\partial_t)^a \nabla^b \phi= \nabla\cdot \{   \nabla_{\d\phi^\sharp}  \partial_t\};
\end{equation}
\begin{equation}
R_\phi{}^\phi = R_{ab} (\partial_\phi)^a \nabla^b \phi = \nabla\cdot \{   \nabla_{\d\phi^\sharp}  \partial_\phi\}.
\end{equation}
Let the indices $A,B\in\{t,\phi\}$ then we can collect these results (four of them altogether) as
\begin{equation}
R_A{}^B = \nabla\cdot \{   \nabla_{(\d x^B)^\sharp}  \partial_{A} \}.
\end{equation}
Unwrapping the covariant derivatives
\begin{equation}
R_{A}{}^{B} = {1\over\sqrt{-g_4}} \, \partial_b \left( \sqrt{-g_4} \, g^{Ba} \, \Gamma^b{}_{aA} \right).
\end{equation}
If we now let the index $i$ range over every coordinate except the two Killing coordinates $t$  and $\phi$, then
\begin{equation}
R_{A}{}^{B} = {1\over\sqrt{-g_4}} \, \partial_i \left( \sqrt{-g_4} \, g^{Ba} \, \Gamma^i{}_{aA} \right).
\end{equation}
Making this all very explicit, there are now four Landau--Lifshitz like results in total.  They are:
\begin{equation}
R_{t}{}^{t} = {1\over\sqrt{-g_4}} \, \partial_i \left( \sqrt{-g_4} \, g^{ta} \, \Gamma^i{}_{at} \right);
\end{equation}
\begin{equation}
R_{t}{}^{\phi} = {1\over\sqrt{-g_4}} \, \partial_i \left( \sqrt{-g_4} \, g^{\phi a} \, \Gamma^i{}_{at} \right);
\end{equation}
\begin{equation}
R_{\phi}{}^{t} = {1\over\sqrt{-g_4}} \, \partial_i \left( \sqrt{-g_4} \, g^{ta} \, \Gamma^i{}_{a\phi} \right);
\end{equation}
\begin{equation}
R_{\phi}{}^{\phi} = {1\over\sqrt{-g_4}} \, \partial_i \left( \sqrt{-g_4} \, g^{\phi a} \, \Gamma^i{}_{a \phi} \right).
\end{equation}
Furthermore, recall that in stationary axisymmetric spacetimes we can always choose coordinates to block diagonalize the metric: $g_{ab} = g_{AB} \oplus g_{ij}$. But then
\begin{eqnarray}
g^{Ba} \, \Gamma^i{}_{aA} &=&  g^{BC} \, \Gamma^i{}_{CA} =  g^{BC} \, g^{ij} \Gamma{}_{jCA} 
\nonumber\\
&=& - {1\over2}  \, g^{BC} \, g^{ij} \; \partial_j g_{CA}.
\end{eqnarray}
So finally we have the relatively compact result
\begin{equation}
R_{A}{}^{B} = -{1\over2}\, {1\over\sqrt{-g_4}} \, \partial_i \left( \sqrt{-g_4} \,  g^{BC} \, g^{ij} \; \partial_j g_{CA} \right).
\end{equation}
This can be rearranged in a number of different ways. As an illustration we point out
\begin{eqnarray}
R_{AB}&=& -{1\over2}\, {1\over\sqrt{-g_4}} \, \partial_i \left( \sqrt{-g_4} \, g^{ij} \; \partial_j g_{AB} \right) 
\nonumber\\
&& + {1\over2} \, g^{ij} \; \partial_i g_{AC} \; g^{CD} \;\partial_j g_{DB}.
\end{eqnarray}
We again see that our two-congruence extension of the Raychaudhuri equation has given us additional useful information regarding the Ricci tensor which might be difficult to extract by other means.

\null
\bigskip

\section{Discussion and conclusions}

In this somewhat technical article we have developed several useful extensions of the usual Raychaudhuri equation.  The main theme has been to relate various linear combinations of components of the Ricci tensor to divergences of suitably defined fluxes.  We have worked with timelike congruences, spacelike congruences, and non-normalized congruences, in all cases being able to say just a little bit more (and sometimes much more) than standard the Raychaudhuri equation would imply. 
One potentially far-reaching result is the ``two congruence'' extension of the Raychauduri equation presented in equations (\ref{E:crossed1}) and (\ref{E:crossed2}). 

{}

\end{document}